\documentclass[preprint,12pt]{elsarticle}
\usepackage{graphicx}
\usepackage{amssymb}
\usepackage{lineno}

\journal{Physica E}

\begin{document}
\begin{frontmatter}

\title{Electron transport in a two-terminal Aharonov-Bohm ring with impurities}

\author{M.A. Kokoreva}
\ead{maria-kokoreva@yandex.ru}

\author{V.A. Margulis}
\ead{theorphysics@mrsu.ru}

\author{M.A. Pyataev}
\ead{pyataevma@math.mrsu.ru}

\address{Mordovian State University 430005, Saransk, Russia}

\begin{abstract}
Electron transport in a two-terminal Aharonov-Bohm ring with a few
short-range scatterers is investigated.
An analytical expression for the conductance as a function
of the electron Fermi energy and magnetic flux
is obtained using the zero-range potential theory.
The dependence of the conductance on positions of scatterers is studied.
We have found that the conductance exhibits
asymmetric Fano resonances at certain energies.
The dependence of the Fano resonances on magnetic field and positions of impurities
is investigated. It is found that collapse of the Fano resonances occurs
and discrete energy levels in the continuous spectrum appear at certain conditions.
An explicit form for the wave function corresponding to the discrete level is obtained.
\end{abstract}


\end{frontmatter}

\begin{linenumbers}

\section*{Introduction}

Quantum ring interferometers of various geometry are intensively studied
theoretically \cite{Gefen84,Buttiker84,Xia92,LiZhang97,Margulis2003,
Voo05,Vargiamidis06,Nakanishi04,Tkachenko00,Bikov00,Pichugin97}
and experimentally
\cite{Bagraev00,Yang00,Wiel00,BikovBakarov00,Yacoby95,RyuCho98,
Kobayashi2002,Kobayashi2004}
for several decades starting from pioneering works \cite{Gefen84,Buttiker84}.
The interest to the systems is stipulated by several new phenomena
such as Aharonov-Bohm oscillations \cite{Pedersen2000,LiuGao1993},
persistent current \cite{Mailly1993,Chandrasekhar91},
electron trapping effect in magnetic field \cite{LiuIsm1993} and so forth.
Recently, the interest to the system has been enhanced due to the experimental detection
\cite{Kobayashi2002,Kobayashi2004} of the Fano resonances \cite{Fano61}
in the electron transport.
These resonances consist of asymmetric peak and dip
on the energy dependence of transmission coefficient.
They are caused by interference  of
bound states and continuum of propagating electron waves.
Fano resonances in electron transport through various systems are
widely studied in literature \cite{KSDK99, KS99E, KSS00, KSRS02,Noc92,CWB01}.

The simplest and most popular model
of quantum interferometer is the one-dimensional ring with two wires attached to it.
This model has been studied in a lot of papers
\cite{Gefen84,Buttiker84,Xia92,LiZhang97,Margulis2003,Voo05,Vargiamidis06}.
In particular, the periodic dependence of conductance on the magnetic field
was obtained in Ref.~\cite{Gefen84} using the scattering matrix approach.
The dependence of the transmission probability on the phase shift
of the transmitted wave for different degrees of coupling
between the current leads and the ring
in the presence of a flux was investigated in Ref.~\cite{Buttiker84}.
It was found that sharp resonances of the Breit--Wigner type exist
in the transmission probability. Two mechanisms giving rise to sharp resonances
were considered.
A one-dimensional quantum waveguide theory for mesoscopic structures
was proposed in Ref.~\cite{Xia92}. In particular, the conductance of two-terminal
Aharonov-Bohm ring as a function of the magnetic flux, the arm lengths,
and the electron wave vector was found.
The resonant-transport properties of the two-terminal ring and
multiply connected ring systems threaded by the magnetic flux
was studied using the tight-binding model \cite{LiZhang97}.
An explicit form for the conductance of the one-dimensional two-terminal
ring threaded by the magnetic flux was found in Ref.~\cite{Margulis2003}
using the zero-range potential theory.
The charge carrier interference in mesoscopic semiconductor rings
formed by two quantum wires in self-organizing silicon quantum wells
was investigated in Ref.~\cite{Bagraev00}.
The periodic dependence of the transmission coefficient and the phase shift
on source-drain bias in the presence of
several $\delta$-shaped barriers in wires was obtained.
The model of one-dimensional quantum ring was used to explain
the results in this work.

The presence of an additional scatterer, for instance a quantum dot or an impurity,
in the interferometer provides new possibilities
to control the electron transport.
Therefore the effect of scatterers on the conductance
is considered in a number of papers
\cite{Voo05,Vargiamidis06,Nakanishi04}.
The electron transport and the persistent current in a mesoscopic ring
connected to current leads was investigated in Ref.~\cite{Vargiamidis06}
using an S-matrix approach. The case of one impurity in
the ring and diametrically opposite contacts was considered.
A contact between wires and ring was defined using
an a priori given energy-independent scattering matrix.
It was shown that in the presence of an impurity
transmission probability of the ring may exhibit resonances
of the Fano type in addition to symmetric Breit-Wigner peaks.
It was found that the transition from the weak to the strong
coupling regime leads to Fano line shapes with gradually smaller
asymmetry parameters, but the positions of the transmission zeros and
ones remain unaffected. The Breit-Wigner like shape is sensitive
to the coupling and disappears in the strong coupling limit.
The presence of an Aharonov-Bohm flux alters the amplitudes of the
Fano resonances, turning them progressively into broad oscillations.
It was found that placing the impurity in a special positions
in the arm and certain values of flux causes
systematic collapse of certain Fano resonances.

However, the effect of impurities on the electron transport
in the quantum ring requires further analysis.
In spite of the large number of theoretical works the case
of many impurities has not been studied in detail yet.
At the same time, strong dependence of the conductance
on positions of impurities might be expected in
the system due to quantum interference phenomena.

The main purpose of the present paper is the investigation
of the electron transport in the Aharonov-Bohm ring containing
several short-range scatterers. The role of scatterers may be
played by impurities or small quantum dots.
We consider the model of one-dimensional ring with two one-dimensional
wires attached to it.
The advantage of the model is the possibility to obtain an analytical
expression for the transmission coefficient of the device.
In the case of one-dimensional wires the one-mode transport regime
takes place and the conductance does not exceed a unit of the conductance
quantum. In this case the transport properties of the device
are determined by the single transmission coefficient.
To obtain this coefficient we use an approach based
on the zero-range potential theory that was used
before in Refs.~\cite{Geiler96,Pavlov00,Geiler98,Exner01,Geyler2003}.
In this method, contacts and impurities are described
by boundary conditions at the points of perturbations.
The general form of the conditions
is defined with the help of the self-adjoint extensions theory
for symmetric operators
\cite{Baz1966,DemkovOstrovsky,PavlovYMN,Albeverio91,Bru03,Krein}.
In the framework of the approach, the system is described by the single Hamiltonian
and parts of the device are connected to each other via energy independent
boundary conditions. At the same time, the scattering matrix
of each contact is energy dependent in this case.
Since the Hamiltonian of the device should be the energy-independent
operator our method seems to be more accurate then the approach
based on a priori defined energy-independent scattering matrix of each contact.

It should be mentioned that non-zero width rings have been studied numerically
\cite{Tkachenko00,Bikov00,Pichugin97}.
Results of those papers are in qualitative increment with the results
obtained in the framework of the one-dimensional model.
In particular, the one-dimensional model describes correctly
quantum interference phenomena if the width of the ring is much smaller
then its radius \cite{Pichugin97}.

\section{Hamiltonian and transmission coefficient}

\begin{figure}[!ht]
\begin{center}
\includegraphics[width=0.8\linewidth]{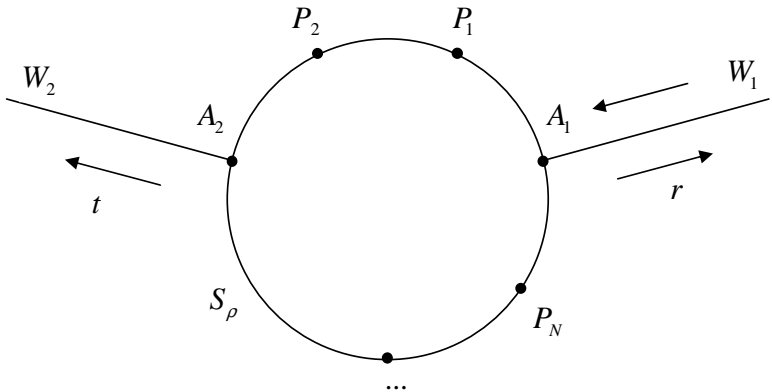}
\caption{\label{Ring}
Two-terminal Aharonov-Bohm ring $\mathbb{S}_\rho$ with impurities
located at points $P_i$. The wires are denoted by $W_1$ and $W_2$.
Points $A_1$ and $A_2$ define positions of contacts, $r$ and $t$
are transmission and reflection amplitudes for the electron wave.}
\end{center}
\end{figure}

Let us consider a quantum ring $\mathbb{S}_\rho$ of radius $\rho$
with two one-dimensional wires $W_1$ and $W_2$ attached to it
at points $A_1$ and $A_2$ respectively.
The scheme of the device is shown in Fig.~\ref{Ring}.

We assume that the ring contains $N$ short-range impurities at points
$P_i$ ($i=1,\ldots N$).
All perturbations are described by the same method using
the boundary conditions for the wave function.
Angles defining positions of perturbations we denote by $\varphi_j$,
where indexes $j=1,2$ correspond to contacts while indexes $j=3,\ldots N+2$
correspond to impurities. It is convenient to use continuous numeration
for all perturbations (contacts and impurities)
since they all are described by the same way.
Without loss of generality we can put $\varphi_1=0$.
The wires $W_j$ are modeled by semiaxes $x\geq 0$.
We suppose that the ring is placed in the magnetic field $B$
perpendicular to the plane of the system. The magnetic flux through
the ring is denoted by $\Phi=\pi \rho^2 B$.

The electron Hamiltonian of the unperturbed ring is given by the following equation
\cite{Margulis2003}
\begin{equation}
H_\rho=\frac{\hbar^2}{2m^*\rho^2}\left(-i\frac{d}{d\varphi}+\eta\right)^2.
\end{equation}
Here $m^*$ is the electron effective mass,
$\eta=\Phi / \Phi_0$ is the number of magnetic flux quanta
and $\Phi_0=2\pi \hbar c / |e|$ is the magnetic flux quantum.

The eigenvalues of the Hamiltonian $H_\rho$ are well known
\begin{equation}
E_m^\eta=\frac{\hbar^2}{2m^*\rho^2}\left(m+\eta\right)^2,
\end{equation}
where $m$ is the magnetic quantum number.

The electron Hamiltonian $H_j$ of each wire $W_j$ has the form
\begin{equation}
 H_j=-\frac{\hbar^2}{2m^*}\frac{d^2}{dx^2}.
\end{equation}

The electron wave function of the device may be represented
in the form of one-column matrix
\begin{equation}
\psi=
\left(\begin{array}{c}
\psi_\rho\\
\psi_1\\
\psi_2\\
\end{array}\right),
\end{equation}
where $\psi_\rho$ is the wave function on the ring  $\mathbb{S}_\rho$,
and $\psi_j$ are wave functions in wires $W_j$.

The impurities are described by the potential
\begin{equation}
V(\varphi)=\sum\limits_{j=3}^{N+2}V_j\delta(\varphi-\varphi_j),
\end{equation}
where $\delta(\varphi)$ is the Dirac $\delta$-function, and
coefficients $V_j$ define the strength of the point perturbations.
It should be mentioned that the potential $V(\varphi)$ is equivalent
to the boundary conditions of the following form
\begin{equation}
\label{imp}
\psi_\rho(\varphi_j)=\frac{1}{v_j}
[\psi'_\rho(\varphi_j+0)-\psi'_\rho(\varphi_j-0)], \quad
j=3,\ldots N+2,
\end{equation}
where  $\psi'_\rho$ means the derivative of $\psi_\rho$ on angle $\varphi$ and
 $v_j=2 m^*\rho^2 V_j/\hbar^2$  is the dimensionless parameter
determining the strength of the point perturbation.

The contacts are modeled with the help of the zero-range potential theory
\cite{Geiler96,Pavlov00,Geiler98,Exner01,Geyler2003,Bru03}.
If there were no contact between different parts
of the system then the Hamiltonian $H_0$ should have the form
\begin{equation}
\label{H0}
H_0=H_\rho\oplus H_1\oplus H_2.
\end{equation}
The direct sum in Eq.~(\ref{H0}) means that each operator
acts on its own part of the wave function and those parts are independent.
Hamiltonian $H$ of the systems is obtained from the Hamiltonian $H_0$
by applying linear boundary conditions for the wave function at points
of contacts.

The most general form of the conditions may be obtained
from the operator extension theory \cite{Bru03}
\begin{eqnarray}
\label{general}
\left\{\begin{array}{l}
\psi_\rho(\varphi_j)=\frac{b_j}{\rho}[\psi'_\rho(\varphi_j+0)-
\psi'_\rho(\varphi_j-0)]+a_j\psi'_j(0),\\
\psi_j(0)=\frac{\bar{a}_j}{\rho}[\psi'_\rho(\varphi_j+0)-
\psi'_\rho(\varphi_j-0)]+c_j\psi'_j(0),\\
\end{array} \right.
\end{eqnarray}
where $\psi'_j$ is the derivative of $\psi_j$ on x,
$a_j$ are complex parameters of dimension of length,
while $b_j$ and $c_j$ are real parameters of the same dimension
(here $j=1,2$).
In the framework of the approach each contact is characterized by
four real parameters. We note that the same general form
of boundary conditions may be obtained from the current conservation
law for each contact.

In the present paper, we will restrict ourselves
by the case of continuous one-dimensional wave function
at the point of contact that corresponds to equal effective width
of the ring and the wires.
One can see from Eqs.~(\ref{general}) that
parameters $a_j$, $b_j$ and $c_j$ should be equal in this case.
It is convenient to represent them in terms of the dimensionless
coupling constant $u_j$ by the equation
\begin{equation}
a_j=b_j=c_j=\rho/u_j.
\end{equation}
Then the boundary conditions are written in the form
\begin{equation}
\label{boundary}
\psi_j(0)=\psi_\rho(\varphi_j) =\frac{1}{u_j}
\{\psi'_\rho(\varphi_j+0)-\psi'_\rho(\varphi_j-0)+\rho\psi'_j(0)\},
\end{equation}
where $j=1,2$.

To obtain the transmission coefficient we have to find a solution of the Schr\"odinger
equation that is the superposition of incident and reflected wave in the first
wire and corresponds to propagated wave in the second wire.
Consequently, the wave function in the wire $W_1$ has the form
\begin{equation}
\label{wavef1}
\psi_1(x)=\exp(-ikx)+r\exp(ikx),
\end{equation}
where $r$ is the reflection amplitude
and $k=\sqrt{2m^*E}/\hbar$ is the electron wave number.
The wave function in the second wire is given by
\begin{equation}
\label{wavef2}
 \psi_2(x)=t\exp(ikx),
\end{equation}
where $t$ is the transmission amplitude.

Since the Hamiltonian $H_0$ is perturbed by the zero-range potentials
the wave-function $\psi_\rho(\varphi)$ can be represented in terms
of the Green function $G_\rho(\varphi,\varphi_j;E)$ of the operator $H_\rho$
\begin{equation}
\label{wavefr} \psi_\rho(\varphi,
E)=\sum_{j=1}^{N+2}A_j(E)G_\rho(\varphi,\varphi_j;E).
\end{equation}
Here $A_j (E)$ are coefficients which should be determined
from the boundary conditions.

The Green function of the Hamiltonian $H_\rho$ is well-known \cite{Margulis2003}
\begin{eqnarray}
\label{fGreen} \nonumber
G_\rho(\varphi,\varphi_j;E)&=&\frac{m^*}{2\hbar^2k}
\left[\frac{\exp\left(i(\varphi_j-\varphi\pm\pi)
(\eta-k\rho)\right)}{\sin{\pi(\eta-k\rho)}}\right.\\
&& \left.-\frac{\exp\left(i(\varphi_j-\varphi\pm\pi)
(\eta+k\rho)\right)}{\sin{\pi(\eta+k\rho)}}\right],
\end{eqnarray}
where ``plus'' sign corresponds to $\varphi \geq \varphi_j$
and ``minus'' sign should be used otherwise.

We denote $Q_{ij}(E)=\frac{\hbar^2}{m^*\rho}G_\rho(\varphi_i,\varphi_j;E)$
and $\alpha_j=\frac{m^*\rho}{\hbar^2}A_j$. The matrix $Q_{ij}(E)$
is called Krein's $Q$-matrix \cite{Bru03,Krein}.
Applying the boundary conditions (\ref{imp}) and (\ref{boundary})
to the wave function given by Eqs.~(\ref{wavef1})--(\ref{wavefr}),
we obtain the system of $N+4$ equations
\begin{eqnarray}
\label{N4eq}
\left\{\begin{array}{l}
\alpha_1Q_{11}+\alpha_2Q_{12}+\ldots+\alpha_{N+2}Q_{1,N+2}=\frac{2\alpha_1+ik\rho(r-1)}{u_1},\\
\alpha_1Q_{21}+\alpha_2Q_{22}+\ldots+\alpha_{N+2}Q_{2,N+2}=\frac{2\alpha_2+ik\rho t}{u_2},\\
\alpha_1Q_{31}+\alpha_2Q_{32}+\ldots+\alpha_{N+2}Q_{3,N+2}=\frac{2\alpha_3}{v_1},\\
\ldots\\
\alpha_1Q_{N+2,1}+\alpha_2Q_{N+2,2}+\ldots+\alpha_{N+2}Q_{N+2,N+2}=\frac{2\alpha_{N+2}}{v_N},\\
1+r=[2\alpha_1+ik\rho(r-1)] /u_1 ,\\
t=[2\alpha_2+ik\rho t]/u_2.
\end{array} \right.
\end{eqnarray}
Using two last equations we can represent $r$ and  $t$ in terms of $\alpha_j$.
Then the system takes the form
\begin{equation}
\label{N2eq}
\sum_{l=1}^{N+2}\left[Q_{jl}-P_j\delta_{jl}\right]\alpha_l=D\delta_{j1},\quad
j=1\ldots,N+2.
\end{equation}
Here we use the following notations
\begin{displaymath}
P_j(E) = \left\{ \begin{array}{ll}
2/(u_j-ik\rho), & j=1,2 \\
2/v_j, & j=3,\ldots, N+2,
\end{array} \right.
\end{displaymath}
\[D(E)=\frac{2ik\rho}{ik\rho-u_1}.\]

The solution of system (\ref{N2eq}) may be represented in the form
\begin{equation}
\label{alpha2}
\alpha_n=\frac{\Delta_n}{\Delta},
\end{equation}
where $\Delta$ is the principal determinant of the system
\begin{equation}
\label{det}
\Delta=\det\left[Q_{jl}-P_j\delta_{jl} \right],
\end{equation}
and $\Delta_n$ is the determinant of the matrix which is obtained from the
basic matrix by replacing of $n$-th column with the column of absolute terms
\begin{equation}
\Delta_n=\det\left[\left(Q_{jl}-P_j\delta_{jl}\right)(1-\delta_{nl})
+ D\delta_{j1}\delta_{nl}\right].
\end{equation}

Taking into account Eq.~(\ref{alpha2}), we can represent the
transmission amplitude in the form
\begin{equation}
\label{amplitude}
t(E)=\frac{2}{u_2-ik\rho}\frac{\Delta_2}{\Delta}.
\end{equation}

We note that Eq.~(\ref{amplitude}) is valid for arbitrary values of
magnetic field and contact parameters. To study the effect of impurity positions
on the electron transport we will restrict ourselves
by the case of equal contacts ($u_1=u_2=u$) and impurities
($v_3=\ldots=v_{N+2}=v$).

\section{Case of one impurity and diametrically opposite contacts}
Let us consider at first the case of one impurity.
Then the system (\ref{N2eq}) consist of three equations.
Using Eq.~(\ref{amplitude}) and the explicit form (\ref{fGreen})
of the Green function $G_\rho(\varphi,\varphi_j;E)$, we obtain the following
equation for transmission amplitude $t$:
\begin{equation}
\label{coeffic}
t(k,\eta)=\frac{F_1(k,\eta)+v
F_2(k,\eta)}{F_3(k,\eta)+v F_4(k)}.
\end{equation}
Here
\begin{equation}
\label{f1}
F_1(k,\eta)=-16ik^3\rho^3\cos{\pi\eta}\sin{\pi k \rho},
\end{equation}
\begin{equation}
\label{f2}
F_2(k,\eta)=8ik^2\rho^2 e^{-i\pi \eta}
\sin[(\pi-\Delta\varphi)k\rho]\sin(\Delta\varphi k\rho),
\end{equation}
\begin{eqnarray}
\label{f3}
\nonumber F_3(k,\eta)&=&4k\rho\{(-u^2+k\rho 2iu+5k^2\rho^2) \sin^2{\pi k\rho}\\
\nonumber &&-4ik\rho(iu+k\rho)\sin{2\pi k\rho}\}\\
&&-16k^3 \rho^3\sin{\pi\eta},
\end{eqnarray}
\begin{eqnarray}
\label{f4}
\nonumber F_4(k)&=&2k\rho(u-ik\rho)\{2\cos{2\pi k\rho}-\cos(2\Delta\varphi k\rho)\\
\nonumber &&-\cos[2(\pi-\Delta\varphi)k\rho]\}\\
\nonumber &&+2(u-ik\rho)^2\sin{\pi k\rho}\cos[(\pi-2\Delta\varphi)k \rho]\\
&&+(5k^2\rho^2+2ik\rho u-u^2)\sin{2\pi k\rho},
\end{eqnarray}
and $\Delta\varphi=\varphi_3-\varphi_2$.

Equation (\ref{coeffic}) coincides with the corresponding equation
for transmission amplitude of the ring without impurities \cite{Margulis2003}
if we take $v=0$ and represent parameter $u$ in terms of the scattering length
$\lambda$ by the equation $u=-2\rho/\lambda$.
As it follows from Eqs.~(\ref{coeffic})--(\ref{f4})
transmission coefficient is a periodic function of the magnetic flux
with the period equal to the flux quantum.
It should be mentioned that the behavior of the transmission coefficient
changes considerably at integer and half-integer values of the magnetic flux $\eta$
since the eigenvalues of the Hamiltonian $H_0$ are degenerated in that case.

\begin{figure}[!ht]
\includegraphics[width=\linewidth]{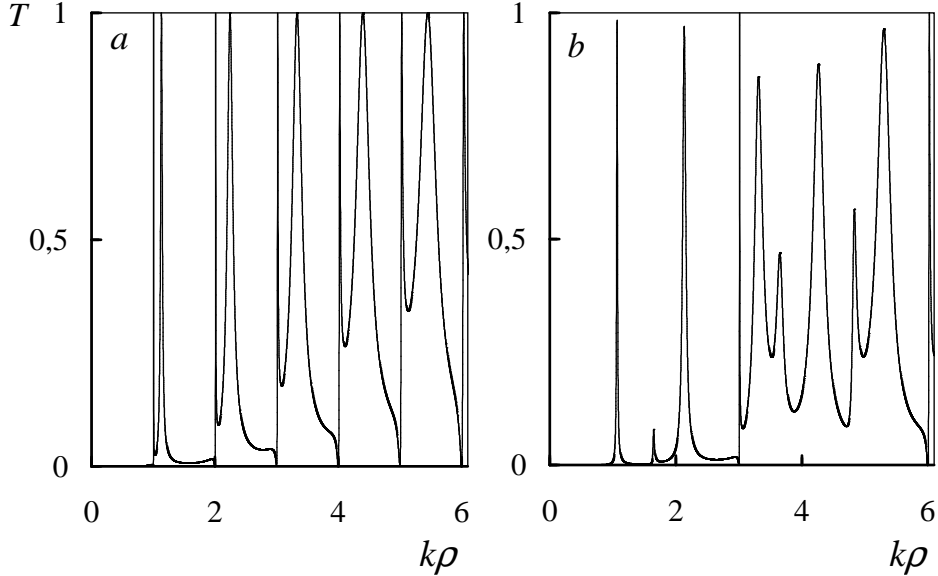}
\caption{\label{f-field014}
Transmission probability $T$ versus the dimensionless wave vector
$k\rho$ in the presence of the magnetic flux $\eta=0.14$.
($a$) The case of the ring without impurities.
($b$) The case of an impurity located at point  $\varphi_3=4\pi/3$.
All figures are plotted for $u=v=10$.}
\end{figure}

We will start with the general case of non-zero magnetic field.
The dependence of the transmission coefficient on the dimensionless parameter
$k\rho$ is shown in Fig.~\ref{f-field014}.
One can see that the transmission coefficient oscillates
as a function of the electron energy.
Oscillations are caused by interference of electron waves
multiply scattered by contacts.
In the absence of the impurity the dependence $T(k\rho)$
exhibits asymmetrical resonances in vicinities of $k\rho=m$ (Fig.~\ref{f-field014}a).
The maximal values of resonance peaks reach unity
while the minimal values of dips are equal to zero.
The impurity leads to destruction of absolute reflection and transmission
that results in decrease of peaks and increase of dip values (Fig.~\ref{f-field014}b).
This result may be explained with help of Eq.~(\ref{coeffic}).
If the impurity is absent ($v=0$) then the numerator in Eq.~(\ref{coeffic})
contains only one term $F_1$ which vanishes at integer values of $k\rho$.
On the other hand, the denominator of Eq.~(\ref{coeffic}) contains
non-trivial real and imaginary parts and does not vanish
at non-integer values of the magnetic flux.
Therefore transmission coefficients has zeros at $k\rho=m$ ($m$ is integer)
if the ring does not contain the impurity.
In presence of the impurity, zeros disappear.
However, some of them are conserved at specific positions of impurity.
In particular, the zero at $k\rho=m$ is conserved
if the following relation holds $\sin{m\Delta\varphi}=0$.

Let us consider the behavior of the transmission coefficient
in vicinities of energy values $E_m^0$ where the asymmetric
Fano resonances can appear.
To study the energy dependence of the transmission coefficient
in the vicinity of $E=E_m^0$ we expend numerator and denominator
of Eq.~(\ref{coeffic}) in the Taylor series up to the first-order terms.
After some simply algebra we obtain the following
asymptotic expression for the transmission amplitude
in the vicinity of $E=E_m^0$
\begin{equation}
\label{t21field}
t(E)\simeq 2i(-1)^m\frac{\pi
m^2\cos{\pi\eta}  \Delta E+ v E_m^0e^{-i\pi\eta}
\sin^2{m\Delta\varphi}} {\pi m(2im -2u-v ) \Delta E+4m E_m^0
\sin^2{\pi\eta}},
\end{equation}
where $\Delta E=E-E_m^0$.

Equation (\ref{t21field}) shows that the transmission coefficient
does not vanish at $E=E_m^0$ in presence of the impurity
if $\sin{m\Delta\varphi}\neq 0$.
However, the zero of transmission at $E=E_m^0$ is conserved if
the position of impurity satisfies the condition $\sin{m\Delta\varphi}=0$
(Fig.~\ref{f-field014}b).
The transmission amplitude near $E=E_m^0$ may be represented in the form
\begin{equation}
\label{t21Fano}
t(E)\simeq \mu_m \frac{E-E_m^0}{E-E_m^{(r)}-i\Gamma_m}.
\end{equation}
Here we use the following notations
\begin{equation}
\label{mu}
 \mu_m=\frac{(-1)^m2im \cos{\pi\eta}}{2im-2u-v},
\end{equation}
\begin{equation}
\label{ER1}
E_m^{(r)}=E_m^0\left(1+\frac{4(2u+v)\sin^2{\pi\eta}}{\pi[(2u+v)^2+4m^2]}\right),
\end{equation}
\begin{equation}
\label{Gamma1}
\Gamma_m=\frac{8mE_m^0\sin^2{\pi\eta}}{\pi[(2u+v)^2+4m^2]}.
\end{equation}
Equation (\ref{t21Fano}) shows that transmission coefficient
has the form of asymmetric Fano resonance \cite{Fano61} in the vicinity of $E_m^0$.

The peak of the Fano resonance corresponds to the
pole $E_m^{(r)}+i\Gamma_m$ on the complex plane.
Here $E_m^{(r)}$ determines the position of the resonance
and $\Gamma_m$ determines the half-width of the resonance curve.
The zeros associated with the Fano resonances are situated
on the real axis at the points $E_m^0$.

One can see from Eq.~(\ref{Gamma1}) that the width $\Gamma_m$ of the Fano
resonance is proportional to $\sin^2{\pi\eta}$ therefore
the width tends to zero as the magnetic flux approaches to
integer values. In this case, the pole $E_m^{(r)}+i\Gamma_m$
and the zero $E_m^0$ on the complex plane coincide and cancel each other
that corresponds to collapse of the Fano resonance.

Let us consider now the dependence of $T$ on $k$ in the case of
integer magnetic flux.
\begin{figure}[!ht]
\includegraphics[width=\linewidth]{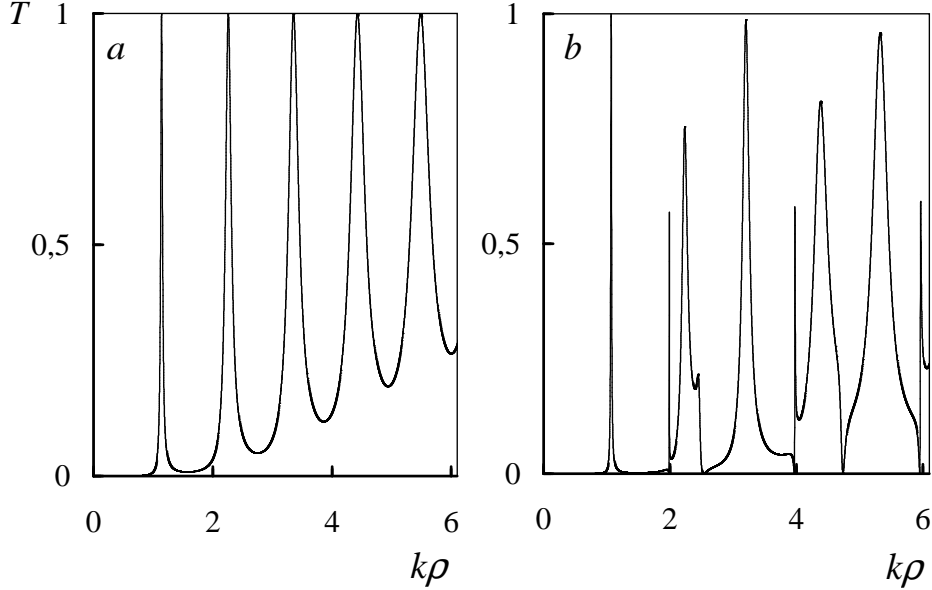}
\caption{\label{f-no_field}
Transmission probability $T$ versus the dimensionless wave vector
$k\rho$ at the zero magnetic field.
($a$) The case of the ring without impurities.
Transmission coefficient has no zeros in this case.
($b$) The case of an impurity located at $\varphi_3=1.475\pi$.
One can see sharp Fano resonances in vicinities of $k\rho=2n$,
where $n$ is integer.}
\end{figure}
The influence of the impurity on the transmission coefficient
is shown in (Fig.~\ref{f-no_field}).
One can see that transmission coefficient has no zeros in
absence of the impurity (Fig.~\ref{f-no_field}a),
as it was shown in previous studies \cite{Margulis2003}.
The presence of the impurity leads to decrease of oscillating maxima and
appearing of Fano resonances (Fig.~\ref{f-no_field}b).

Zeros associated with the Fano resonances does not coincide with the values $E_m^0$.
We denote by $\kappa_m$ the values of the electron wave vector
corresponding to the zeros of transmission coefficient.
According to equations (\ref{f1}) and (\ref{f2}) values of $\kappa_m$
are determined by the equation
\begin{equation}
\label{kappa}
2\kappa\rho\sin({\pi\kappa\rho})
-v\sin[(\pi-\Delta\varphi)\kappa\rho]\sin(\Delta\varphi\kappa\rho)=0.
\end{equation}
The following approximate expression for $\kappa_m$ may be obtained
if $|\sin(n\Delta\varphi)|\ll 1$:
\begin{equation}
\label{deltak}
\kappa_n \approx \frac{n}{\rho}- \frac{v\sin^2{n\Delta\varphi}}{2\pi n
\rho},
\end{equation}
The corresponding electron energy is given by
\begin{equation}
\label{E0}
E_n^{(z)}=\frac{\hbar^2\kappa_n^2}{2m^*}\simeq\frac{\hbar^2}{2m^*\rho^2}
\left(n-\frac{v\sin^2{n\Delta\varphi}}{2\pi n}\right)^2.
\end{equation}
Let us consider the behavior of the transmission amplitude
in the vicinity of $E_m^{(z)}$. For this purpose we
expand the numerator and the denominator of $t(k)$ in Eq.~(\ref{coeffic})
in the Taylor series near $k_m=m/\rho$ neglecting the second order
terms with regard to $\sin{m\Delta\varphi}$.
Then we get the following approximate equation for $t(k)$:
\begin{equation}
\label{t21eta0}
t(k,0)\simeq\frac{2i\pi(-1)^{m+1} m^2\rho(k-\kappa_m)}
{\pi m \rho(v-2i)(k-k_m)+v(u-im)\sin^2{m\Delta\varphi}}.
\end{equation}
Now we introduce the following notations:
\begin{equation}
\label{mu21}
\mu_m=-\frac{2i(-1)^m m}{(v-2i)},
\end{equation}
\begin{equation}
\label{ReFano2}
E_m^{(r)}\simeq E_m^0\left(1-\frac{2v(vu+2m)
\sin^2{m\Delta\varphi}}{\pi m^2(v^2+4)}\right)
\end{equation}
and
\begin{equation}
\label{ImFano2}
\Gamma_m\simeq \frac{2E_m^0 v(vm-2u)\sin^2{m\Delta\varphi}}{\pi m^2(v^2+4)}.
\end{equation}
Then we obtain the following equation for the transmission amplitude in the vicinity of
$\kappa_m$
\begin{equation}
\label{t21Fano2}
t(E)\simeq \mu_m \frac{E-E_m^{(z)}}{E-E_m^{(r)}-i\Gamma_m}.
\end{equation}

\begin{figure}[!ht]
\includegraphics[width=\linewidth]{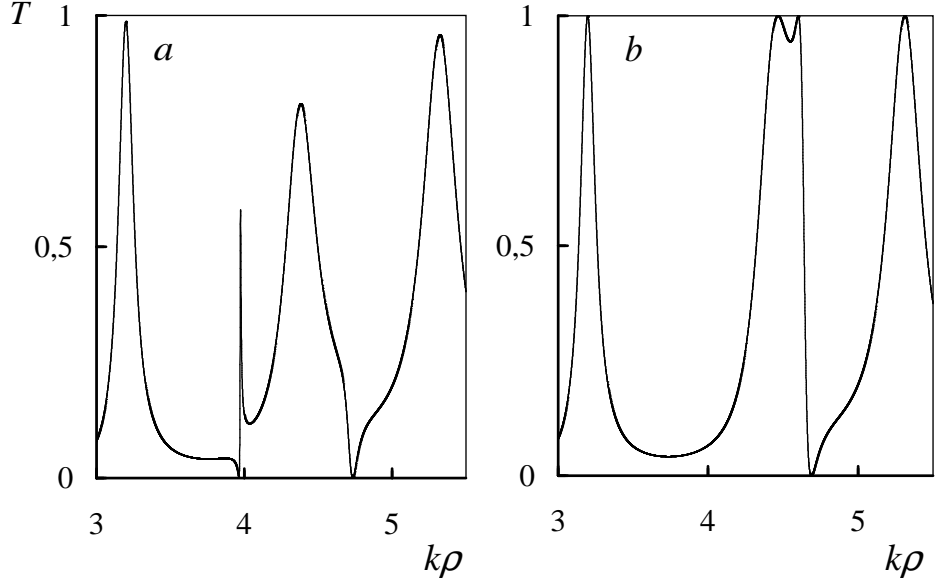}
\caption{\label{f-no_field2}
Transmission probability $T$ versus the dimensionless wave vector
$k\rho$ at the zero magnetic field.
($a$) The impurity is located at point $\varphi_3=1.475\pi$.
One can see the Fano resonance in the vicinity of $k\rho=4$.
($b$) Collapse of the Fano resonance occurs if
the impurity is located at $\varphi_3=1.5\pi$.}
\end{figure}

One can see from Eq.~(\ref{t21Fano2}) that the transmission amplitude
has the form of the Fano resonance in the vicinity of $E_m^{(z)}$.
The the width of the resonance curve is determined by the
position of the impurity.
If the position satisfies the condition $\sin{m\Delta\varphi}=0$
then the collapse of the Fano resonance
in the vicinity of $E_m^{(r)}$ occurs (Fig.~\ref{f-no_field2}).
It should be mentioned that the similar result was obtained in Ref.~\cite{Vargiamidis06}
in the framework of the scattering matrix approach.

We note that the Fano resonance near $E_m^0$ disappear if the magnetic field is absent
and the impurity position is determined by the equation $\varphi_3=l\pi/n$ where
$n$ is the integer divisor of $m$.
In the case of the non-zero magnetic field the resonances at those $E_m^0$
appear again.

Let us consider in detail the case of half-integer magnetic flux.
The transmission coefficient vanishes for all electron energies
if there is no impurity on the ring and contacts are located
in diametrically opposite points.
The presence of the impurity leads to appearance of the non-zero
transmission as follows from Eq.~(\ref{coeffic}).
The phenomenon might be explained in terms of the electron wave phase.
The electron waves spreading on the ring from the point $A_1$
in different directions get phase shifts $\pi(k\rho+\eta)$
and $\pi(k\rho-\eta)$ at the point $A_2$ (Fig.~\ref{Ring}).
The phase difference is equal to $2\pi\eta$
therefore those waves have opposite phases in the second lead
and cancels each other for all energies
at half-integer values of the magnetic flux $\eta$.
Consequently the reflection coefficient equals to unity and
the device may be considered as an ideal electron mirror.
The impurity breaks destructive interference
and the non-zero transmission coefficient appears.

\begin{figure}[!ht]
\includegraphics[width=\linewidth]{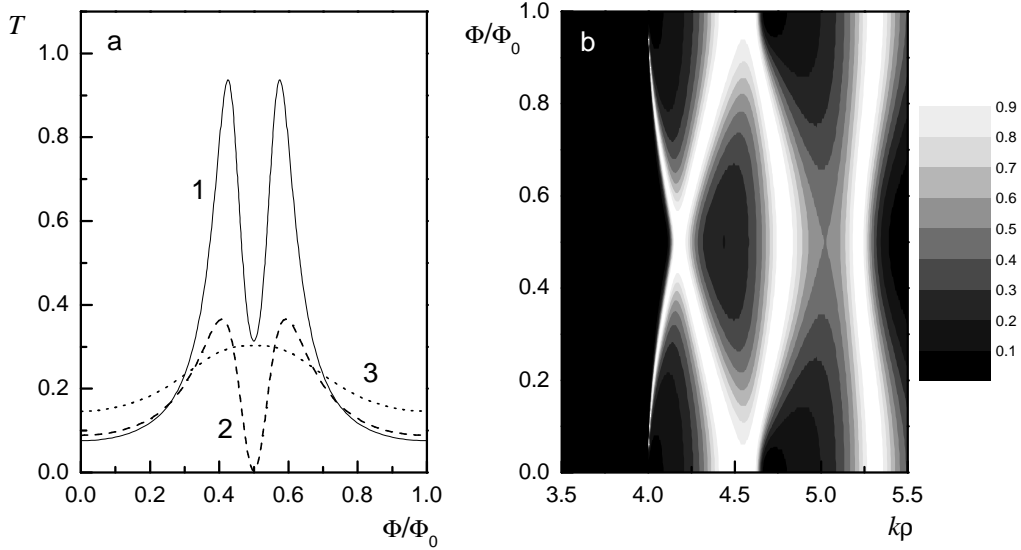}
\caption{\label{f-vsfield}
($a$) Transmission probability $T$ versus the magnetic flux $\eta$
for the case of diametrically opposite contacts for several
impurity positions:  $\varphi_3=1.02\pi$ (solid line 1),
$\varphi_3=1.033\pi$ (dashed line 2)
and $\varphi_3=1.06\pi$ (dotted line 3).
($b$) Transmission probability $T$ as a function of magnetic flux $\eta$
and dimensionless wave vector $kr$ for the case of
diametrically opposite contacts. The impurity is located
at the point $\varphi_3=3\pi/2$. }
\end{figure}

Let us consider the dependence of the transmission coefficient on
the magnetic field that is represented in Fig.~\ref{f-vsfield}.
It follows from Eq.~(\ref{coeffic}) that $T(k,\eta)$
is a periodic function of $\eta$ with the period equal to $1$.
The dependence of the transmission coefficient on $\eta$ contains one or two maxima
on the period (Fig.~\ref{f-vsfield}a)
depending on the electron energy and the impurity position.
Figure~\ref{f-vsfield}b shows that maxima of $T(\eta)$
can conjugate and diverge with variation of the electron energy.
The similar features are present on the dependence of the
transmission coefficient on the magnetic field and the impurity position.

\section{Non-opposite position of contacts}

Let us consider now the system with contacts attached at arbitrary
points ($\varphi_2- \varphi_1\neq \pi$).
The transmission amplitude in this case is given by
\begin{equation}
\label{coeffic2}
t(k,\eta)=\frac{\tilde{F}_1(k,\eta)+
v\tilde{F}_2(k,\eta)}{\tilde{F}_3(k,\eta)+
v\tilde{F}_4(k)},
\end{equation}
where
\begin{eqnarray}
\label{tf1}
\nonumber \tilde{F}_1(k,\eta)&=&-8ik^3\rho^3e^{-i\varphi_2\eta}
\{\sin[(2\pi-\varphi_2)k\rho]\\
&&+e^{2i\pi\eta}\sin(\varphi_2 k\rho)\},
\end{eqnarray}
\begin{eqnarray}
\label{tf2}
\tilde{F}_2(k,\eta)=8ik^2\rho^2 e^{-i\varphi_2\eta}
\sin[(2\pi-\varphi_3) k\rho]\sin(\Delta\varphi k\rho),
\end{eqnarray}
\begin{eqnarray}
\label{tf3}
\nonumber \tilde{F}_3(k,\eta)&=& 2k\rho\{u^2\cos{2\pi k\rho}\\
\nonumber &&+(iu+k\rho)^2\cos{[2(\pi-\varphi_2)k\rho]}\\
\nonumber &&+k\rho[4k\rho\cos{2\pi\eta}-(2iu+5k\rho)\cos{2\pi k\rho}\\
&&+4(u-ik\rho)\sin{2\pi k\rho}]\},
\end{eqnarray}
and
\begin{eqnarray}
\label{f4n}
\nonumber
\tilde{F}_4(k)&=&2ik\rho(iu+k\rho)
\{\cos[2(\pi-\Delta\varphi)k\rho]+\\
\nonumber
&&+\cos{[2(\pi-\varphi_3)k\rho]}-2\cos{2\pi k\rho}\}+\\
\nonumber
&&+(iu+k\rho)^2\{\sin{[2(\pi-\varphi_3)k\rho]}-\\
\nonumber
&&-\sin{[2(\pi-\varphi_2)k\rho]}-\sin{[2(\pi-\Delta\varphi)k\rho]}\}+
\\&&
+(5k^2\rho^2+2ik\rho u-u^2)\sin{2\pi k\rho}.
\end{eqnarray}

If the ring has no impurities and the magnetic field is absent
then Eq.~(\ref{coeffic2}) takes the form
\begin{equation}
t(k,0)=\tilde{F}_1(k,0)/\tilde{F}_3(k,0),
\end{equation}
where $\tilde{F}_1(k,0)$ is given by
\begin{equation}
\label{F2n2}
\tilde{F}_1(k,0)=-16ik^3\rho^3\sin{k\pi\rho}\cos{[(\pi-\varphi_2)k\rho]}.
\end{equation}
Equation (\ref{F2n2}) shows that the transmission coefficient
vanishes if $\sin{k\pi\rho}=0$ or $\cos{[(\pi-\varphi_2)k\rho]=0}$
and the function $\tilde{F}_3$ remains non-zero.

\begin{figure}[!ht]
\includegraphics[width=\linewidth]{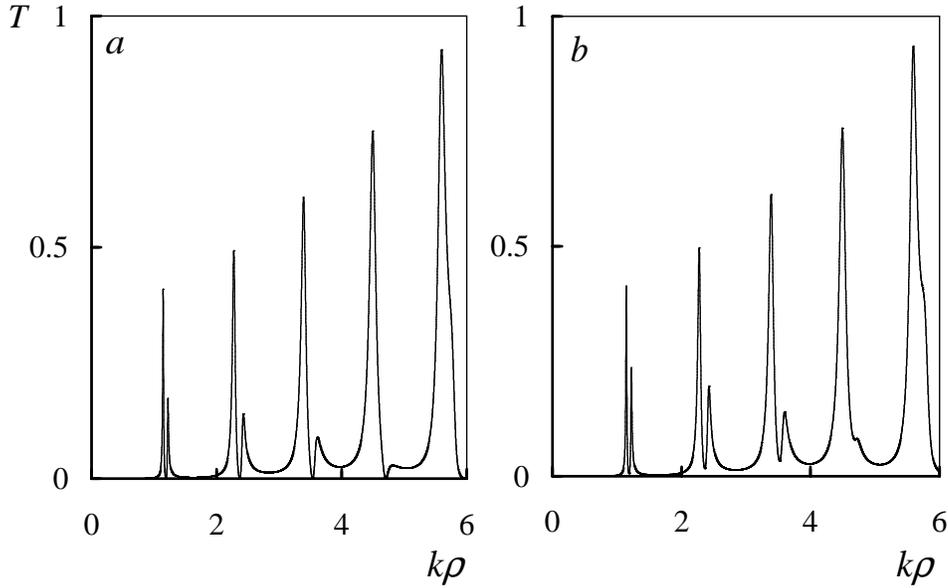}
\caption{\label{f-nd}
Transmission probability $T$ versus the dimensionless wave vector
$k\rho$ for the case of non-opposite contacts
($\varphi_1=0$, $\varphi_2=14\pi/15$).
The impurity is located at the point $\varphi_3=1.1\pi$.
($a$) The magnetic field is absent.
($b$) The magnetic flux is equal to $\eta=0.05$.}
\end{figure}

The presence of the impurity leads to shift of transmission zeros
from the values $k\rho=n$ (Fig.~\ref{f-nd} a).
The positions of zeros are determined by the equation
\begin{eqnarray}
\label{eq_zeros}
\nonumber &&2k\rho\sin{k\pi\rho}\cos{[(\pi-\varphi_2)k\rho]}\\
&&-v\sin[(2\pi-\varphi_3)k\rho]\sin(\Delta\varphi k\rho)=0.
\end{eqnarray}

Equation (\ref{eq_zeros}) shows that if the impurity
position is defined by $\Delta\varphi=\pm l\pi/m$ ($l$ and $m$ are integer)
then zeros are situated at $k\rho=n$ where $n$ is integer multiply $m$.
The magnetic field
leads to disappearing of zeros at $k\rho=n$ (Fig.~\ref{f-nd} b).
In contrast to the collapse of the Fano resonances
the depth of dips decreases while the width remains finite.
Transmission zeros on the complex plane are shifted from the real axis
and corresponding dips do not reach zero
in the case of the applied magnetic field.
The energy determining the position
of the zero on the complex plane acquire an imaginary part
which increases with the  magnetic field.
Thus there are two different mechanisms of
disappearing of zeros in the system:
the first one is collapse of the Fano resonance and
the second one is shift of the transmission zero
from the real axis in the complex plane.
We mention that the collapse of the Fano resonance
is accompanied by increase in symmetry of the system
while the second mechanism is caused by decrease in symmetry.

We mention that the transmission reaches maximal value
when the system has mirror symmetry.
In particular, the height of the conductance peak is maximal
if the impurity is located in the middle of
an arc connecting contacts and parameters of contacts are equal.

\section{Discrete levels in the continuum}

It is known \cite{KSDK99, KS99E, KSS00, KSRS02}
that the collapse of the Fano resonance
is accompanied by appearance of the discrete level in the continuous spectrum.
This level corresponds to localized state on the ring.
The wave function $\psi^\mathrm{d}_\rho$ corresponding to this state
has to satisfy the boundary conditions (\ref{imp}) and  (\ref{boundary})
and vanish at the points of contacts
\begin{equation}
\label{discrete}
\psi^\mathrm{d}_\rho(\varphi_1)=\psi^\mathrm{d}_\rho(\varphi_2)=0.
\end{equation}

If the electron energy coincides with the eigenvalue $E_m^\eta$ of the Hamiltonian $H_\rho$
then the wave function is a linear combination of the corresponding eigenfunctions.
The eigenvalues $E_m^\eta$ are non-degenerated if $2\eta$ is not integer. In this case, the eigenfunction
has the form
\begin{equation}
\label{discrete1}
\psi_\rho^{m}(\varphi)=\frac{1}{\sqrt{2\pi}}e^{im\varphi}.
\end{equation}
The function $\psi_\rho^{m}(\varphi)$ has no zeros
and therefore it does not satisfy the condition (\ref{discrete}).

At integer and half-integer values of the magnetic flux $\eta$ the
eigenvalues $E_m^\eta$ are double-degenerated.
The general form of the wave function for these cases is given by
\begin{equation}
\label{d_psi}
\psi_\rho(\varphi)=
C_1e^{im\varphi}+C_2 e^{-i(m+2\eta)\varphi},
\end{equation}
where $C_1$ and $C_2$ are some coefficients.
It is easy to obtain the function of the form (\ref{d_psi})
which vanishes at $\varphi_1=0$
\begin{equation}
\label{d_psi1}
\psi_\rho^\mathrm{d}(\varphi)=
\frac{1}{\sqrt{\pi}}\exp\left(-i\eta\varphi\right)
\sin\left[\left(m+\eta\right)\varphi\right].
\end{equation}
According to Eq.~(\ref{discrete}) the function $\psi_\rho^\mathrm{d}(\varphi)$
has to vanish at the point $\varphi_2$.
Since the wave function (\ref{d_psi1}) is smooth
it has to vanish at the point $\varphi_3$
to satisfy the boundary condition (\ref{imp}).
Therefore the discrete level appears in the
continuous spectrum if the wave function given by Eq.~(\ref{d_psi1}) has  zeros
at all perturbation points on the ring.
Positions of the impurity and the second contact corresponding
to the collapse of the Fano resonance
at $E_m^\eta$ are given by the following equations
\begin{equation}
\label{phi_2_3}
\varphi_2=\frac{n_2}{m+\eta}\pi,\quad \varphi_3=\frac{n_3}{m+\eta}\pi.
\end{equation}
where $n_2$ and $n_3$ are arbitrary integer numbers.
Equation (\ref{phi_2_3}) coincides with the condition
of the collapse of the Fano resonance obtained in previous sections.

In the case of $E\neq E_m^\eta$ the wave function corresponding to the discrete level
should have the form
\begin{equation}
\label{d_psi2}
\psi^\mathrm{d}_m(\varphi)=A_3G(\varphi,\varphi_3,E).
\end{equation}
Applying the boundary condition (\ref{imp}) to the function (\ref{d_psi2}) we obtain
the equation
\begin{equation}
\label{dis_E}
Q_{33}(E)+\frac{2}{v}=0.
\end{equation}
determining the energy of the discrete level associated with the impurity.
In the case of the zero magnetic field Eq.~(\ref{dis_E}) is written in the form
\begin{equation}
\cot \pi k\rho+\frac{2k\rho}{v}=0.
\end{equation}
However Eq.~(\ref{dis_E}) is not enough for appearance of the discrete level
in the case of attached leads. The condition (\ref{discrete})
has to be satisfied in addition to
Eq.~(\ref{dis_E}). That means the contacts should be situated
at zeros of the function (\ref{d_psi2}).
Applying the boundary conditions (\ref{discrete})
to the wave function (\ref{d_psi2}) we obtain the conditions of appearance
of the discrete level
\begin{equation}
Q_{31}(E)=Q_{32}(E)=0.
\end{equation}
This equation is valid only at special positions of the impurity
and at special energies.
We note that the function (\ref{d_psi2}) has no zeros at irrational values
of the magnetic flux $\eta$.
Furthermore the position of zeros depends on the strength of the impurity potential.
Therefore the positions of contacts have to be coordinated
with the value of impurity potential.
The coordination does not take place for all levels at the same time.
Hence only one discrete level
can appear in the continuous spectrum at $E\neq E_m^\eta$.
By virtue of the symmetrical property of the function (\ref{d_psi2}),
the appearance of this level is more likely
if the contacts are situated on equal distance from the impurity.

\section{Case of two impurities}

In the general case of arbitrary location of impurities and contacts,
the dependence of transmission coefficient on the electron energy contains
different type of overlapping resonances and oscillations.
The dependence is irregular in this case and hardly analyzable,
therefore we will restrict ourselves by the configuration of diametrically
opposite contacts.
\begin{figure}[!ht]
\includegraphics[width=\linewidth]{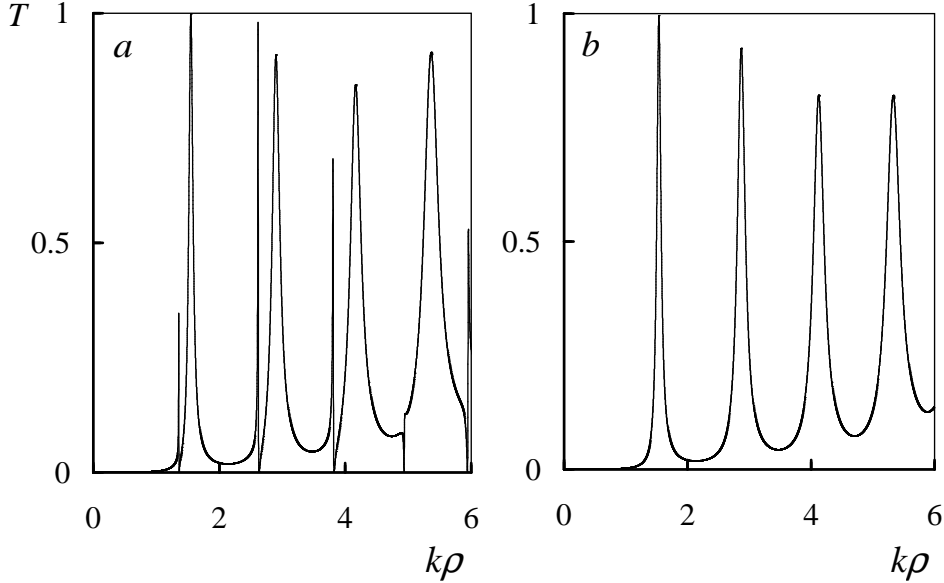}
\caption{\label{fig2PFano}
Transmission probability $T$ versus the dimensionless wave vector
$k\rho$ for the case of diametrically opposite contacts and the zero magnetic field
at $u=10$ and $v=5$.
($a$) Two impurities are located at points  $\varphi_3=0.8\pi$ and $\varphi_4=1.15\pi$.
($b$) The impurities are located on equal distance from each contact
($\varphi_3=0.83\pi$ and $\varphi_4=1.17\pi$).}
\end{figure}

If the angles defining positions of the impurities are incommensurable
and magnetic field is absent then the dependence $T(k\rho)$
exhibits Fano resonances (Fig.~\ref{fig2PFano}a) in the vicinity of the
values $k\rho=m$ where $m$ is integer.
If the impurities are located at equal distances from the contacts
($\varphi_4=2\pi-\varphi_3$) then the simultaneous collapse of
all Fano resonances occurs (Fig~\ref{fig2PFano}b).

\begin{figure}[!ht]
\includegraphics[width=\linewidth]{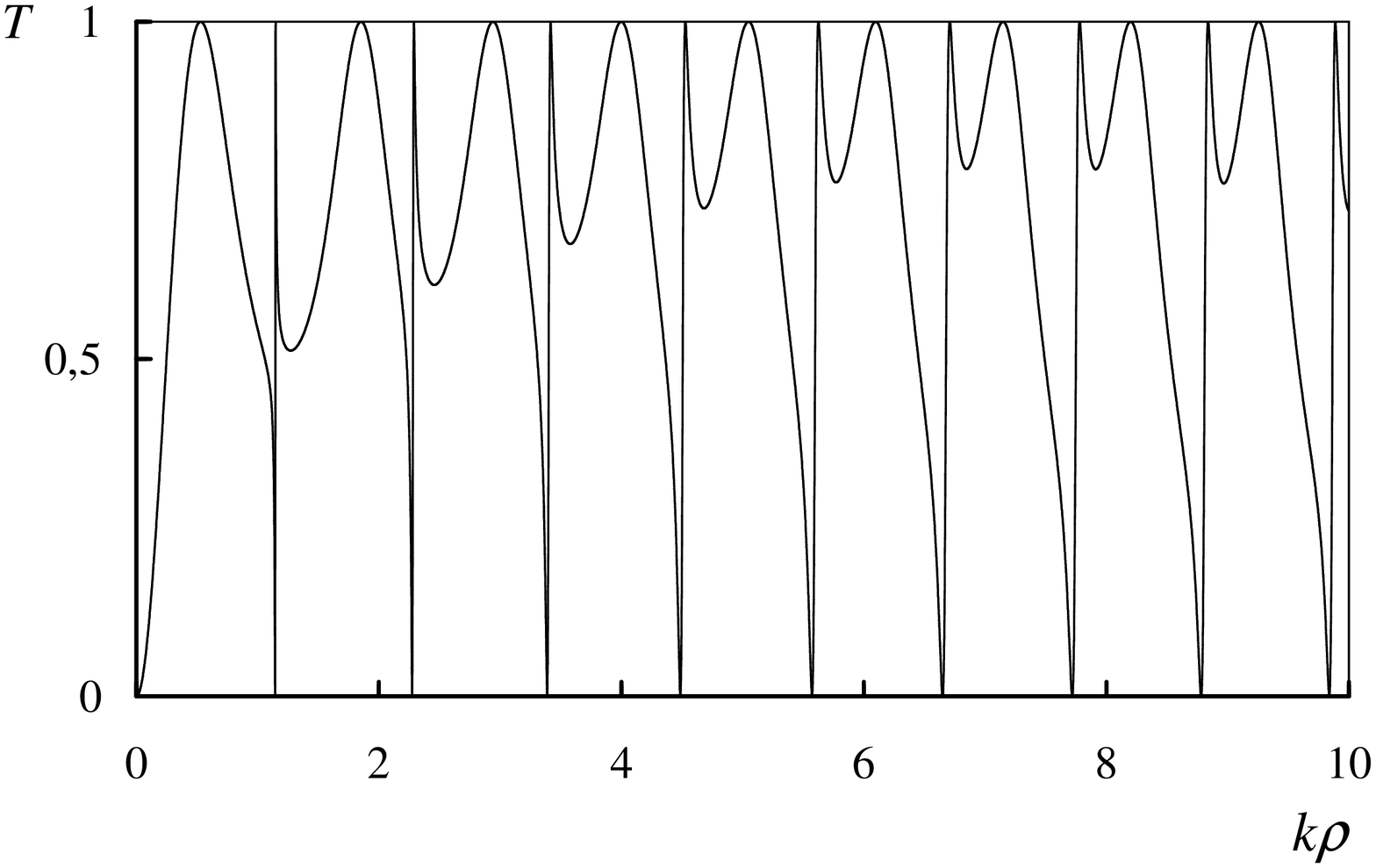}
\caption{\label{P2pi1516pi15}
Transmission probability $T$ versus the dimensionless wave vector
$k\rho$ for the case of diametrically opposite contacts and zero magnetic field
at $u=10$ and $v=5$ in presence of two impurities located diametrically opposite at points
$\varphi_3=\pi/15$ and $\varphi_4=16\pi/15$.}
\end{figure}
The wave function of the discrete level
associated with the collapse
in the case of two impurities at $E\neq E_m^\eta$ has the form
\begin{equation}
\label{d_psi3}
\psi^\mathrm{d}_m(\varphi)=A_3G(\varphi,\varphi_3,E)+A_4G(\varphi,\varphi_4,E).
\end{equation}
The energy of the discrete level is defined by the equation
\begin{equation}
\label{dis_E2}
\left|
\begin{array}{cc}
Q_{33}(E)+\frac{2}{v}&Q_{34}(E)\\
Q_{43}(E)&Q_{44}(E)+\frac{2}{v}
\end{array}
\right|=0.
\end{equation}
The condition (\ref{discrete}) should be satisfied in addition to Eq.~(\ref{dis_E2}).
At zero magnetic field, the Green function $G(\varphi,\varphi_j,E)$ has the property
\begin{equation}
\label{G_even}
G(\varphi_j-\Delta \varphi,\varphi_j,E)=G(\varphi_j+\Delta \varphi,\varphi_j,E).
\end{equation}
Taking into account Eq.~(\ref{G_even}), one can see that
the wave function
\begin{equation}
\label{d_psi4}
\psi^\mathrm{d}_m(\varphi)=A_3[G(\varphi,\varphi_3,E)-G(\varphi,\varphi_4,E)].
\end{equation}
satisfies the boundary condition (\ref{discrete}) at arbitrary energy.
Therefore the presence of the mirror symmetry in the system
leads to  simultaneous collapse of all Fano resonances.
The Fano resonances appear again if the symmetry is
broken by shift of any impurity or
by different values of point potentials ($v_1 \neq v_2$).
Collapse of several Fano resonances in vicinities of $k\rho=m$
occurs if the positions of the impurities
are given by $\varphi_3=n_3\pi/m$ and $\varphi_4=n_4\pi/m$
where $n_3$, $n_4$ and $m$ are integer.

If the impurities are located in diametrically opposite points
($\varphi_4=\varphi_3+\pi$)
then the maximal values of transmission peaks reach unity (Fig.~\ref{P2pi1516pi15})
in contrast to the case of single impurity.
Hence the addition of the second impurity at the diametrically opposite
point leads to increase of the conductance due do constructive interference.

It should be mentioned that the destructive interference at
half-integer values of the magnetic flux leads to the perfect reflection
if the system has mirror or inverse symmetry. Therefore
the transmission coefficient vanishes at the half-integer magnetic flux
if two identical impurities are located at diametrically opposite points
($\varphi_4=\varphi_3+\pi$) or at the same distance from contacts
($\varphi_4=2\pi-\varphi_3$).

Zeros of transmission coefficient disappear in the magnetic field
with magnetic flux $\eta\neq n/2$ where $n$ is integer.
The similar effect has been considered in the case of one impurity
and illustrated by Fig.~\ref{f-nd}.

\section*{Conclusion}

We have investigated the electron transport in the two-terminal
Aharonov--Bohm ring with impurities.
Using the zero-range potential theory we have obtained analytical
expressions for the electron transmission coefficient
as a function of the electron energy.
The effect of the magnetic field and position of impurities
on the electron transport has been studied.
We have found that the dependence of the transmission coefficient
on the electron energy exhibits oscillations caused by the interference of
electron waves on the ring.
The presence of impurities can lead either to decrease
or to increase in the transmission coefficient.
In particular, the transmission increases in presence of impurities
due to breaking of the destructive interference if the value magnetic flux
through the ring equals to half-integer number of the magnetic flux quanta.
The ring without impurities may be considered as ideal electron mirror
in this case.

Our analysis shows that the dependence of the transmission
coefficient on the electron energy exhibits Fano resonances at certain energies.
These resonances consist of the sharp transmission peak and
the nearby transmission zero.
In the complex plane of energy, the Fano resonance is represented
by pole and nearby zero of the transmission amplitude.
Resonances arise as a result of interaction between discrete
level and continuous spectrum. The necessary condition of
appearing of the Fano resonance in the quantum ring is partial
symmetry breaking by the magnetic field or the asymmetrical
location of impurities or contacts.
We have found that the collapse of Fano resonances
occurs at certain parameters of the system.
In this case, the pole and the zero of the transmission amplitude coincide
and cancel each other.
The discrete level immersed in the continuous spectrum arises
in this case.
In particular, collapse of the Fano resonance in the vicinity of the
discrete level $E_m^\eta$ occurs if
all perturbations on the ring are situated at zeros of the
wave function defined by Eq.~(\ref{d_psi1}).

Another mechanism of disappearing of transmission zeros is possible in the
system in addition to the collapse of Fano resonances.
The zero may be shifted from the real axis in the complex plane.
In this case, the dip of transmission coefficient is conserved but
the minimum value does not reach zero.
We mention that the collapse of the Fano resonance is accompanied by
increase in symmetry of the system while the shift of the zero
is accompanied by decrease in symmetry.

The work is supported by Russian funding program
"Development of Scientific Potential of Higher School"
(Grant No 2.1.1/2656).

\end{linenumbers}

\end{document}